\documentclass[10pt,twocolumn,letterpaper,twocolumn]{article}
\usepackage{ol2}
\usepackage[draft]{hyperref}
\usepackage{amsmath,amsfonts,amssymb}
\usepackage{calc}
\usepackage{multirow}
\usepackage{graphicx}
\usepackage{bm}
\usepackage{color}
\usepackage{epsfig}
\usepackage{ifthen}

\def\be{ \begin{equation} }
\def\ee{ \end{equation} }
\def\bea{ \begin{eqnarray} }
\def\eea{ \end{eqnarray} }
\def\bse{ \begin{subequations} }
\def\ese{ \end{subequations} }

\def\i{\,\text{i}}
\def\e{\,\text{e}}
\def\const{\,\text{const}}

\def\to{\rightarrow}

\def\d{\text{d}}
\def\U{\mathbf{U}}

\def\H{\mathbf{H}}
\def\c{\mathbf{c}}

\def\i{{\rm{i}}}
\def\f{{\rm{f}}}
\def\phase{\phi}

\def\area{\mathcal{A}}

\def\fwhm{\xi}

\begin{document}

\twocolumn[
\title{High-fidelity local addressing of trapped ions and atoms by composite sequences of laser pulses}

\author{Svetoslav S. Ivanov$^{1,2}$ and Nikolay V. Vitanov$^{1}$}

\address{
$^1$Department of Physics, Sofia University, 5 James Bourchier Blvd, 1164 Sofia, Bulgaria \\
$^2$Institut Carnot de Bourgogne, Universit\'e de Bourgogne, 9, Av A. Savary, 21078 Dijon, France
}

\begin{abstract}
A vital requirement for a quantum computer is the ability to locally address, with high fidelity, any of its qubits without affecting their neighbors.
We propose an addressing method using composite sequences of laser pulses, which reduces dramatically the addressing error in a lattice of closely spaced atoms or ions,
 and at the same time significantly enhances the robustness of qubit manipulations.
To this end, we design novel high-fidelity composite pulses for the most important single-qubit operations.
In principle, this method allows one to beat the diffraction limit, for only atoms situated in a small spatial region around the center of the laser beam are excited, well within the laser beam waist.
\end{abstract}

]

Scalable quantum computers depend critically on the ability to perform local addressing of their individual qubits \cite{NC}.
In a Paul ion trap, which is one of the most promising scalable platforms for the future quantum computer \cite{roadmap},
 local addressing is the ability to operate on a single ion by focused laser light while keeping the neighboring ions unaffected.
When the number of ions increases, the distance between them diminishes and local addressing becomes one of the principal experimental challenges.
For example, in a recent experimental demonstration of the Toffoli gate \cite{Blatt2009} most of the error was attributed to addressing error, as the neighboring ions were seeing 7\% of the central Rabi frequency.

In this Letter, we propose a method for high-fidelity local addressing applicable to various types of atomic qubits: trapped ions,
atoms in  optical lattices, quantum dots, etc. To this end, we present new narrowband (NB) and passband (PB) composite pulses\footnote{We follow the usual NMR terminology, which is related to the features of the excitation profile, rather than the radiation.}, which are specially designed for local addressing.
The excitation profiles of such pulses allow one to manipulate only a single qubit, as the outer parts of the spatial
laser beam profile practically do not excite its neighbors, although the latter may be subjected to significant laser intensity.
Moreover, with a PB pulse one enhances the robustness of qubit manipulations, thereby eliminating errors due to imperfectly calibrated and fluctuating laser intensity and laser beam pointing instability.

The technique of composite pulses was introduced in nuclear magnetic resonance (NMR) \cite{spinecho, NMR, Wimperis}
 as a powerful tool for manipulation of spins by magnetic fields.
A composite pulse compensates the imperfections of a single pulse, which is the traditional tool used to drive a quantum transition,
 and it consists of a sequence of pulses, each with a well-defined phase.
The composite phases are determined from the conditions imposed on the desired overall excitation profile.
In particular, in a NB composite pulse only the qubits seeing pulse areas within a narrow range around some value $\area$ are subjected to transformation, while qubits seeing areas outside this range remain unaffected in the end of the composite sequence.

Most known composite pulses, however, are inappropriate for local addressing due to the sidebands and the slowly vanishing tails in their excitation profiles.
The few composite pulses that allow local addressing \cite{Haffner} are very sensitive to variations in the pulse area.
We present here a systematic method, which allows us to construct high-fidelity composite pulses, which eliminate these drawbacks and which can be made robust and accurate to any desired order.

A two-state quantum system $\psi_1\leftrightarrow\psi_2$, subjected to an external coherent electromagnetic field, obeys the Schr\"{o}dinger equation, $\i \hbar\partial_t \mathbf{c}(t) = \H(t)\mathbf{c}(t)$.
Here the vector $\c(t) = [c_1(t), c_2(t)]^T$ contains the two probability amplitudes and the Hamiltonian is
$\H(t) = (\hbar/2) \Omega(t) \e^{-\i D(t)} |\psi_1\rangle\langle \psi_2| + \text{h.c.}$,
with $D(t)=\int_{t_\i}^{t}\Delta (t^{\prime})\d t^{\prime}$, where $\Delta=\omega_0 -\omega$ is the detuning between the laser carrier frequency $\omega$ and the Bohr transition frequency $\omega_0$, and $\Omega(t)$ is the Rabi frequency.
The amplitudes at the end of the interaction $\c(t_\f)$ are obtained from the initial ones with the propagator $\U$: $\c(t_\f)=\U(t_\f,t_\i)\c(t_\i)$, which can be parameterized with the complex Cayley-Klein parameters $a$ and $b$ (obeying $|a|^2+|b|^2=1$),
\be
\U = \left[\begin{array}{cc}  a  & b \\  -b^{\ast} & a^{\ast} \end{array} \right].
\ee
For resonance ($\Delta=0$), the Schr\"{o}dinger equation
has an exact solution regardless of the shape of $\Omega(t)$
and the parameters $a$ and $b$ are determined only
by the pulse area $A=\int_{t_\i}^{t_\f}\Omega(t)\d t$: $a=\cos(A/2) $, $b=-\i\sin(A/2)$. The transition probability is $p = |b|^2=\sin^2\left( A/2 \right)$.
A constant phase shift $\phase$ in the driving field, $\Omega(t)\to\Omega(t)\e^{\i\phase}$, is mapped onto the propagator as
\be
\U_\phase = \left[ \begin{array}{cc} a & b \e^{-\i\phase} \\ -b^{\ast}\e^{\i\phase}  & a^{\ast} \end{array}\right].
\ee
\begin{table}[tb]
\centering
\begin{tabular}{|l|c|l|} \hline
\multicolumn{3}{|c|}{Narrowband sequences}\\
\hline
 $N_{2n+1}(\area)$ & $A$ & phases $(\phi_2;\phi_3;\phi_4;\ldots;\phi_{n+1})$ \\
\hline
N$_5(\pi)$                   & $\pi$    & (0.839; 1.420)\\
N$_9(\pi)$                   & $\pi$    & (0.426; 1.490; 0.858; 1.300) \\
N$_{13}(\pi)$                & $\pi$    & (1.103; 0.876; 0.154; 1.708; \\
                             & & \quad 1.020; 0.229) \\
N$_{21}(\pi)$                & $\pi$    & (1.073; 0.919; 0.131; 1.831; \\
                             & & \quad 1.156; 0.721; 0.096; 1.521; \\
                             & & \quad 0.812; 1.954) \\
N$_7(\frac{\pi}{2})$         & $3\pi/7$ & (0.471; 1.196; 1.315) \\
N$_7(\frac{\pi}{\sqrt{2}})$  & $\pi/2$  & (0.577; 1.161; 1.573) \\
N$_7(\frac{\pi}{2\sqrt{2}})$ & $3\pi/8$ & (1.532; 0.800; 0.698) \\
N$_7(\sqrt{2}\pi)$           & $3\pi/4$ & (1.505; 0.823; 0.609) \\
\hline \hline
\multicolumn{3}{|c|}{Passband sequences}\\ \hline
 $P_{2n+1}(\area)$   & $A$      & phases $(\phi_2;\phi_3;\phi_4;\ldots;\phi_{n+1})$ \\
\hline
P$_7(\pi)$                   & $\pi$    & (0.508; 1.337; 1.083) \\
P$_{17}(\pi)$                & $\pi$    & (1.235; 0.721; 0.934; 0.126; \\
                             & & \quad 1.872; 1.515; 0.873; 0.217) \\
P$_9(\frac{\pi}{2})$         & $3\pi/5$ & (1.270; 1.106; 0.464; 0.053) \\
P$_{17}(\frac{\pi}{2})$          & $2\pi/3$ & (0.459; 0.097; 0.302; 1.445; \\
                             & & \quad 0.829; 1.324; 1.290; 0.995) \\
P$_9(\frac{\pi}{\sqrt{2}})$  & $3\pi/5$ & (0.676; 0.87; 1.503; 1.836) \\
P$_9(\frac{\pi}{2\sqrt{2}})$ & $3\pi/5$ & (0.356; 1.517; 0.957; 1.023) \\
P$_9(\sqrt{2}\pi)$           & $3\pi/5$ & (1.909; 1.197; 0.861; 0.660) \\
\hline
\end{tabular}
\caption{Phases $\phi_k$ (in units $\pi$) for some NB (N$_N$) and PB (P$_N$) sequences of $N=2n+1$ phased resonant pulses of area $A$: $A_0 A_{\phi_2}A_{\phi_3}\cdots A_{\phi_{n+1}}\cdots A_{\phi_3}A_{\phi_2}A_0$. We set $n_1=n$, $n_2=0$ for all NB pulses $N_{2n+1}(\area)$; $n_1=2$, $n_2=1$ for $P_7(\area)$ and $P_9(\area)$; $n_1=6$, $n_2=3$ for $P_{17}(\pi)$; $n_1=4$, $n_2=3$ for $P_{17}(\pi/2)$. For all sequences we set $n_3=0$. For $\area\neq\pi$ we also impose Eq. \eqref{eq1}. The pulse area is $A=\pi$ for $N_5(\pi)$, $N_9(\pi)$, $N_{13}(\pi)$, $N_{21}(\pi)$, $P_7(\pi)$ and $P_{17}(\pi)$; $A=3\pi/7$, $\pi/2$, $3\pi/8$ and $3\pi/4$, respectively, for $N_7(\pi/2)$, $N_7(\pi/\sqrt{2})$, $N_7(\pi/(2\sqrt{2}))$ and $N_7(\sqrt{2}\pi)$; $A=2\pi/3$ for $P_{17}(\pi/2)$; $A=3\pi/5$ for $P_9(\pi/2)$, $P_9(\pi/\sqrt{2})$, $P_9(\pi/(2\sqrt{2}))$ and $P_9(\sqrt{2}\pi)$.
 }
\label{table1}
\end{table}

We assume for simplicity, and possible experimental convenience, that all pulse areas are equal\footnote{Allowing for different pulse areas, and thereby seemingly for more free parameters, does not seem to be advantageous. We have found through extensive examination that the total pulse area (and the pulse duration) of the composite sequence is not reduced.}. A sequence of $N$ pulses $A_{\phase_k}$, each with area $A$ and phase $\phase_k$, produces the propagator
\be\label{U^N}
\U^{(N)} = \U(A_{\phase_{N}}) \U(A_{\phase_{N-1}}) \cdots \U(A_{\phase_{1}}).
\ee
We consider here an odd number of pulses, $N=2n+1$, although this restriction is not essential.
We consider composite sequences, which are symmetric with respect to reversal of pulses, i.e. the phases should obey $\phase_{k}=\phase_{N+1-k}$ $(k=1,2,\dots,n)$;
 this ``anagram'' condition annuls the imaginary part of the propagator element $U^{(N)}_{11}$.
Since the overall phase of the sequence is irrelevant, but only the relative phases of the pulses matter for the dynamics, we set $\phase_1=\phase_{N}=0$;
 hence we are left with $n$ independent phases, which are treated as free parameters.
For NB pulses we require transition probability $p\approx 0$ for pulse areas in the vicinity of $A=0$ (a flat bottom); for PB pulses we also require $p\approx \const$ around the desired overall area $\area$ (a flat plateau).
Thus, the phases for a NB sequence are derived from the conditions
\bse\label{conditions}
\begin{align}
&[U_{11}^{(N)}]_{A=\area}=\cos(\area/2), \label{eq1}\\
&[\partial^k_A U_{11}^{(N)}]_{A=0} = 0 \quad (k=2,4,\dots,2n_1), \label{eq2}
\end{align}
\ese
with $n_1=n-1$ and $\partial^k_A\equiv\partial^k / \partial A^k$.
For a $n_2$-fold flat-top PB sequence we add the conditions 
\be\label{conditions2}
[\partial^k_A U_{11}^{(N)}]_{A=\area} = 0\quad (k=1,2,\dots,n_2),
\ee
with $n_1+n_2+1=n$.
For a sequence with a target phase $\varphi$ we must also have
\be
[\arg U_{21}^{(N)}]_{A=\area}=\varphi.
\label{additional1}
\ee
This phase can be stabilized with PB pulses, for which
\be
[\arg \partial^k_A U_{21}^{(N)}]_{A=\area}=0\quad (k=1,2,\dots,n_3).
\label{additional2}
\ee
Thus a phase-stable PB sequence consists of $n=n_1+n_2+n_3+2$ pulses.
For target area $\area=\pi$, Eqs.~\eqref{eq1} and \eqref{conditions2} for even $k$ and \eqref{additional2} for odd $k$ are fulfilled identically.

\begin{figure}[tb]
\includegraphics[width=0.95\columnwidth]{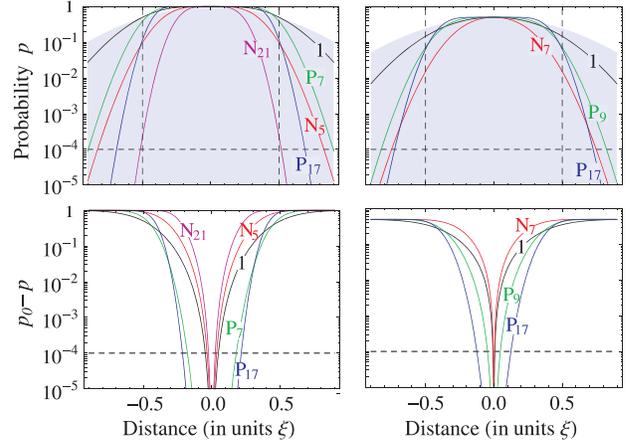}
\caption{
\emph{Top frames:} Excitation probability $p$ in a laser field with a Gaussian spatial profile (grey-shaded) with FWHM of Rabi frequency $\fwhm$,
 vs the distance from its center for various composite pulses from Table \ref{table1}:
N$_5(\pi)$, P$_7(\pi)$, N$_{21}(\pi)$ and P$_{17}(\pi)$ (left) and N$_7(\pi/2)$, P$_9(\pi/2)$ and P$_{17}(\pi/2)$ (right).
The excitation profiles of single $\pi$ and $\pi/2$ pulses are shown too.
\emph{Bottom frames:} Deviation $p_0-p$, with $p_0=1$ (left) and $p_0=0.5$ (right) being the desired excitation probability.
}
\label{fig1}
\end{figure}

Using the method described above, we have constructed new NB and PB composite pulses;
examples are presented in Table \ref{table1} for the most often used areas.
Figure \ref{fig1} illustrates the excitation profiles for different composite pulses of target area $\pi$ (left frames) and $\pi/2$ (right frames).
At the center of the laser spot (origin) the probability of excitation is {$p_0=\sin^2(\area/2)$}, whereas at the wings it naturally decreases.
The logarithmic scale allows us to examine the fidelity of the profile against the $10^{-4}$ quantum information benchmark \cite{NC}.
We assume that the laser beam is a spot with a full-width-at-half-maximum (FWHM) of intensity $\xi/\sqrt{2}$;
this implies FWHM of Rabi frequency $\xi$ for single-photon transitions.
Remarkably, suppression of unwanted neighbor excitation below the $10^{-4}$ benchmark is achieved
at up to 15\% of the peak Rabi frequency (corresponding to distance $0.83\xi$ from the center of the spot) with the N$_5$ pulse,
27\% (distance $0.70\xi$) with P$_{17}$, and  48\% (distance $0.51\xi$) with N$_{21}$.
In this manner we can beat the diffraction limit as the excitation is localized in a spatial range, which, with sufficiently many ingredient pulses, can be made smaller than the beam waist.
The physical reason for this suppression is the destructive interference of the ingredient pulses in the composite sequence.

The bottom frames of Fig.~\ref{fig1} show the infidelity of the target qubit itself
 and reveal that PB composite pulses can greatly enhance the robustness of manipulation of the target qubit without losing selectivity.
For target area $\area=\pi$ (left frame) an infidelity of $10^{-4}$ is encountered at offset $0.05\fwhm$ for a single pulse,
 while the admissible offset reaches $0.18\fwhm$ for P$_7(\pi)$ and $0.21\fwhm$ for P$_{17}(\pi)$.

\begin{table}[tb]
\begin{center}
\begin{tabular}{|l|c|l|}
\hline
 Sequences & $A$ & phases $(\phi_2;\phi_3;\phi_4;\ldots;\phi_{n+1})$ \\
\hline
N$_7(\pi,\frac{3\pi}{2})$                      & $\pi$    & (1.256; 0.792; 0.072) \\
P$_{11}(\pi,\frac{3\pi}{2})$                   & $\pi$    & (0.221; 1.109; 0.753; 1.304; \\
 & & \quad 1.878) \\
N$_9(\frac{\pi}{2},\frac{\pi}{2})$             & $3\pi/4$ & (1.074; 0.935; 0.173; 1.562) \\
P$_{13}(\frac{\pi}{2},\frac{\pi}{2})$          & $3\pi/4$ & (0.959; 1.048; 0.367; 1.967; \\
 & & \quad 1.511; 0.860) \\
N$_9(\frac{\pi}{\sqrt{2}},\frac{3\pi}{2})$     & $3\pi/5$ & (1.326; 0.958; 0.137; 0.791) \\
P$_{13}(\frac{\pi}{\sqrt{2}},\frac{3\pi}{2})$  & $3\pi/5$ & (0.183; 0.978; 1.421; 0.769; \\
 & & \quad 1.924; 1.916) \\
\hline
\end{tabular}
\end{center}
\caption{Phases $\phi_k$ (in units $\pi$) for some NB (N$_N(\area,\varphi)$) and PB (P$_N(\area,\varphi)$) sequences of $N=2n+1$ phased resonant pulses of area $A$: $A_0 A_{\phi_2}A_{\phi_3}\cdots A_{\phi_{n+1}}\cdots A_{\phi_3}A_{\phi_2}A_0$, which produce phased rotations of angle $\varphi$ at area $\area$.
We set $n_2=n_3=0$ for all NB pulses $N_{2n+1}(\area,\varphi)$. We have $n_2=1$, $n_3=4$ for $P_{11}(\area,\varphi)$; $n_2=n_3=1$ for $P_{13}(\area,\varphi)$. For all sequences we set $n_1=2$ and we impose Eq. \eqref{additional1}. For $\area\neq\pi$ we also impose Eq. \eqref{eq1}. The pulse area is $A=\pi$ for $N_7(\pi,3\pi/2)$ and $P_{11}(\pi,3\pi/2)$; $A=3\pi/4$ for $N_9(\pi/2,\pi/2)$ and $P_{13}(\pi/2,\pi/2)$; $A=3\pi/5$ for $N_9(\pi/\sqrt{2},3\pi/2)$ and $P_{13}(\pi/\sqrt{2},3\pi/2)$.
}
\label{table2}
\end{table}

\begin{figure}[tb]
\centering
\includegraphics[width=0.80\columnwidth]{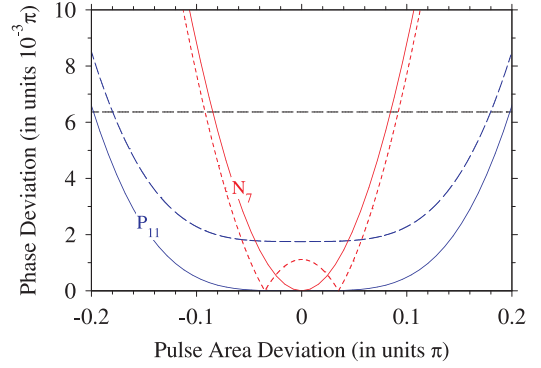}
\caption{
Phase deviation $\left|\phi-3\pi/2\right|$ vs pulse area deviation of the composite pulses $N_7(\pi,\frac{3\pi}{2})$ and $P_{11}(\pi,\frac{3\pi}{2})$, shown in Table \ref{table2}. The dashed curves are for frequency mismatch $\Delta=0.001/T$. The horizontal line is the threshold above which the infidelity exceeds $10^{-4}$.}
\label{fig2}
\end{figure}

Our scheme can be further exploited, in conjunction with Eqs. \eqref{additional1} and \eqref{additional2}, to generate various NB and PB sequences of specified target phases $\varphi$ at the expense of additional pulses.
Table \ref{table2} lists a set of composite sequences, which produce experimentally relevant phased rotations.
Larger sequences achieve stabilization of the phase $\varphi$.
Figure \ref{fig2} illustrates the robustness of the composite phase $\varphi=3\pi/2$ \cite{Blatt2009}
produced by two of our phase-stabilized sequences from Table \ref{table2}.
Remarkably, one can perform high-fidelity local addressing even with pulse area deviation of 20\%.
In addition, we have found that pulse area noise with relative amplitude 5\% introduces an absolute error in the target phase of $2.5\times 10^{-3}\pi$.

To conclude, the new composite sequences designed for high-fidelity local addressing in a lattice of closely spaced qubits, are of potential application to Paul ion traps \cite{ions,Haffner} and ultracold atoms in optical lattices \cite{Bloch}.
They allow one to stabilize both the rotation angle and the phase of the desired qubit rotation.
This technique can be adapted to addressing on the vibrational sidebands, which should allow to construct high-fidelity two-qubit operations.

This work has been supported by the European Commission project FASTQUAST and the Bulgarian NSF grants VU-I-301/07, D002-90/08 and DMU02-19/09.


\begin{thebibliography}

\bibitem{NC} M. A. Nielsen and I. L. Chuang, \emph{Quantum Computation and Quantum Information} (Cambridge University Press, UK, 2000).

\bibitem{roadmap} \htmladdnormallink{http://qist.lanl.gov/qcomp\_map.shtml}{http://qist.lanl.gov/qcomp\_map.shtml}

\bibitem{Blatt2009} T. Monz, K. Kim, W. H\"{u}nsel, M. Riebe, A. S. Villar, P. Schindler, M. Chwalla, M. Hennrich, and R. Blatt, Phys. Rev. Lett. \textbf{102}, 040501 (2009).

\bibitem{spinecho} E. L. Hahn, Phys. Rev. \textbf{80}, 580 (1950).

\bibitem{NMR}
M.H. Levitt and R. Freeman, J. Magn. Reson. \textbf{33}, 473 (1979); 
R. Freeman, S. P. Kempsell, and M. H. Levitt, J. Magn. Reson. \textbf{38}, 453 (1980); 
H. M. Cho, R. Tycko, A. Pines, and J. Guckenheimer, Phys. Rev. Lett. \textbf{56}, 1905 (1986); 
M.H. Levitt, Prog. NMR Spectrosc. \textbf{18}, 61 (1986);
R. Freeman, \emph{Spin Choreography} (Spektrum, Oxford, 1997).

\bibitem{Wimperis}{S. Wimperis, J. Magn. Reson. \textbf{109}, 221 (1994).}

\bibitem{Haffner}
H. H\"{a}ffner, C.F. Roos, R. Blatt, Phys. Rep. \textbf{469}, 155 (2008).

\bibitem{ions}
S. Gulde, M. Riebe, G. P. T. Lancaster, C. Becher, J. Eschner, H. H\"{a}ffner, F. Schmidt-Kaler, I. L. Chuang and R. Blatt, Nature \textbf{421}, 48 (2003); 
F. Schmidt-Kaler, H. H\"{a}ffner, M. Riebe, S. Gulde, G. P. T. Lancaster, T. Deuschle, C. Becher, C. F. Roos, J. Eschner, and R. Blatt, Nature \textbf{422}, 408 (2003); 
N. Timoney, V. Elman, S. Glaser, C. Weiss, M. Johanning, W. Neuhauser, and C. Wunderlich, Phys. Rev. A \textbf{77}, 052334 (2008).

\bibitem{Bloch}
I. Bloch, T. W. H\"{a}nsch, and T. Esslinger, Nature \textbf{403}, 166 (2000);
I. Bloch, J. Dalibard, and W. Zwerger, Rev. Mod. Phys. \textbf{80}, 885 (2008); 
C. Weitenberg, M. Endres, J. F. Sherson, M. Cheneau, P. Schau\ss, T. Fukuhara, I. Bloch, and S. Kuhr, arXiv:1101.2076.

\end{thebibliography}
\end{document}